\documentclass[twocolumn,prc,showpacs,showkeys,preprintnumbers,amsmath,amssymb,floatfix]{revtex4}

\usepackage{graphicx}
\usepackage{dcolumn}
\usepackage{bm}

\newcommand{\re}{\text{Re}}
\newcommand{\im}{\text{Im}}

\newcommand{\PDG}{Yao:2006px}

\begin{document}

\preprint{}

\title{Effective $\bar{K}N$ interaction based on chiral SU(3) dynamics}

\author{Tetsuo~Hyodo}
\email{thyodo@ph.tum.de}
\affiliation{%
Physik-Department, Technische Universit\"at M\"unchen, 
D-85747 Garching, Germany
}%

\affiliation{%
Yukawa Institute for Theoretical Physics,
Kyoto University, Kyoto 606--8502, Japan
}%
 
\author{Wolfram Weise}
\affiliation{%
Physik-Department, Technische Universit\"at M\"unchen, 
D-85747 Garching, Germany 
}%
 
\date{\today}
\begin{abstract}
    The effective $\bar{K}N$ interaction based on chiral SU(3) 
    coupled-channel dynamics is derived and its extrapolation below the
    $\bar{K}N$ threshold is studied in detail. Starting from the 
    coupled-channel scattering equations, we eliminate the channels other 
    than $\bar{K}N$ and obtain an effective interaction in the single 
    $\bar{K}N$ channel. An equivalent local potential in coordinate space is
    constructed such as to reproduce the full scattering amplitude of the 
    chiral SU(3) coupled-channel framework. We discuss several realistic 
    chiral SU(3)-based models in comparison to reach conclusions about the 
    uncertainties involved. It turns out that, in the region relevant to the
    discussion of deeply bound $\bar{K}$-nuclear few-body systems, the 
    resulting energy-dependent, equivalent local potential is substantially 
    less attractive than the one suggested in previous purely 
    phenomenological treatments. 
\end{abstract}




\maketitle

\section{Introduction}\label{sec:intro}

The quest for quasibound antikaon-nuclear states has become a persistently 
hot topic in nuclear physics. It is argued that if the $\bar{K}N$ 
interaction is sufficiently strong and attractive so that $\bar{K}$-nuclear 
bound systems can be formed, with binding energies so large that they fall 
below the $\bar{K}N\rightarrow \pi\Sigma$ threshold, such states could be 
narrow. An experiment performed at KEK with stopped $K^-$ on 
$^4$He~\cite{Suzuki:2004ep} seemed to indicate such deeply bound narrow 
structures. However, the repetition of this experiment with better 
statistics~\cite{Sato:2007sb} did not confirm the previously published 
results. The FINUDA measurements with stopped $K^-$ on $^{6,7}$Li and 
$^{12}$C targets ~\cite{Agnello:2005qj} suggested an interpretation in terms
of quasibound $K^-pp$ clusters with binding energy $B(K^-pp) = (115 \pm 9)$ 
MeV and width $\Gamma = (67\pm 16)$ MeV. However, this interpretation was 
subsequently criticized in Refs.~\cite{Oset:2005sn,Magas:2006fn} with the 
argument that the observed spectrum may be explained by final-state 
interactions of the produced $\Lambda p$ pairs. Although these issues are 
unsettled, the experimental search for kaonic nuclei continues vigorously.

Calculations of strong binding of antikaons in a nuclear medium based on 
chiral SU(3) dynamics have a long history (see, \textit{e.g.},
Refs.~\cite{Brown:1993yv,Waas:1996fy,Waas:1997pe,Waas:1997tw,Lutz:2001dq,
Lutz:2007bh}), starting from the early discussions of kaon condensation in 
dense matter~\cite{Kaplan:1986yq,Kaplan:1987sc}. The recent revival of this 
theme was prompted by Akaishi and Yamazaki~\cite{Akaishi:2002bg,Dote:2003ac,
Dote:2004wd}, who used a simple potential model [unconstrained by chiral 
SU(3)] to calculate bound states of few-body systems such as $K^-pp$, 
$K^-ppn$, and $K^-pnn$. The possibility that such systems could be highly 
compressed was suggested in Ref.~\cite{Dote:2003ac}, but this was 
considered~\cite{Weise:2007rd} to be an artifact of an unrealistic 
nucleon-nucleon interaction being used. 

Faddeev calculations were performed for $\bar{K}NN$ with phenomenological 
input~\cite{Shevchenko:2006xy,Shevchenko:2007ke}, and with a leading-order 
chiral interaction~\cite{Ikeda:2007nz}. Both studies found a $K^-pp$ 
quasibound state above the $\pi\Sigma N$ threshold with relatively large 
width. The $\bar{K}NN$ system has also been studied in 
Ref.~\cite{Yamazaki:2007cs} by using a variational approach with 
phenomenological local potentials~\cite{Akaishi:2002bg}, leading to a bound 
state at about 50 MeV below the $\bar{K}NN$ threshold.

One should note that the predictive power of all such investigations is 
limited because the energy range of the $\bar{K}N$ interaction relevant for 
deeply bound kaonic nuclei lies far below the $\bar{K}N$ threshold. 
Constraints from $\bar{K}N$ scattering and from kaonic hydrogen measurements
are restricted to $\sqrt{s} \ge 1432$ MeV. The only experimental information
available below the $\bar{K}N$ threshold is the invariant mass spectrum in 
the $\pi\Sigma$ channel where the $\Lambda(1405)$ resonance is observed. 
However, fitting the interaction to these data involves a subtlety as we 
will discuss in this paper. It turns out, namely, that the peak position in 
the $\pi\Sigma$ mass spectrum is not to be identified with the pole position
of the $\Lambda(1405)$ in the $\bar{K}N$ amplitude. Thus, naively assigning 
a $K^- p$ binding energy of 27 MeV to the $\Lambda(1405)$, as is frequently 
done in phenomenological potential approaches, is not justified in view of 
the strong $\bar{K}N \leftrightarrow \pi\Sigma$ coupled-channel dynamics. In
any event, we need to extrapolate the interaction calibrated around the 
$\bar{K}N$ threshold down to much lower energies to explore the possible 
existence of deeply bound $\bar{K}$-nuclear systems. Ambiguities in 
performing such extrapolations certainly arise and require a careful and 
detailed assessment.

A reliable and realistic starting point for a theory of low-energy 
$\bar{K}N$ interactions is the coupled-channel approach based on the chiral 
SU(3) meson-baryon effective Lagrangian, developed and first applied in 
Ref. \cite{Kaiser:1995eg}, and subsequently expanded by several groups 
\cite{Oset:1998it,Oller:2000fj,Lutz:2001yb}. Unitarization of the chiral 
interaction correctly reproduces the $\bar{K}N$ scattering observables and 
provides a framework for generating the $\Lambda(1405)$ resonance 
dynamically as a $\bar{K}N$ quasibound state embedded in the strongly 
interacting $\pi\Sigma$ continuum. Given that this approach is successful 
over a  wide range of energies and a variety of channels, we would like to 
investigate in detail what chiral SU(3) dynamics tells us about the 
$\bar{K}N$ interaction below threshold.

For variational calculations of few-body systems involving antikaons, one 
must use a realistic effective $\bar{K}N$ interaction, preferentially in the
form of a potential. This potential is generally complex and energy 
dependent. It must be constrained to reproduce the scattering amplitudes in 
vacuum, and it must encode the full coupled-channel dynamics. The first 
attempts in this direction, which use a schematic effective interaction, 
have been reported in Refs.~\cite{Weise:2007rd,Dote:2007rk}. Here we would 
like to explicitly derive such an effective interaction in the single 
$\bar{K}N$ channel and construct an equivalent, energy-dependent local 
potential, starting from chiral SU(3) coupled-channel scattering. This 
interaction can then be used in $\bar{K}$-nuclear few-body 
calculations~\cite{Dotesan}.

This paper is organized as follows. In Sec.~\ref{sec:formulation} we 
introduce the chiral coupled-channel framework for $S=-1$ meson-baryon 
scattering. We present a general framework for constructing an effective 
interaction with reduced number of channels in a system of coupled-channel 
scattering equations, with full incorporation of the dynamics in the 
eliminated channels. In Sec.~\ref{sec:effective} this formalism will be 
applied to the $I=0$ $\bar{K}N$ channel, showing how the $\pi\Sigma$ and 
other channels affect the $\bar{K}N$ single-channel interaction. We study 
the pole structure of the $\Lambda(1405)$ in the complex energy plane and 
discuss the physical origin of the singularities. We also construct the 
$I=1$ effective interaction and estimate theoretical uncertainties for
subthreshold extrapolations in both the $I=0$ and $I=1$ $\bar{K}N$ channels.
Finally we derive, for practical use, an ``equivalent'' local potential in 
coordinate space in Sec.~\ref{sec:local}. A comparison is performed with 
amplitudes calculated from the phenomenological potential of 
Ref.~\cite{Yamazaki:2007cs}, and substantial differences are pointed out. 
The last section presents a summary and conclusions.

\section{Formal Framework}\label{sec:formulation}

\subsection{Chiral SU(3) dynamics with coupled channels}\label{subsec:ChU}

Consider meson-baryon scattering in the strangeness $S=-1$ channel. The 
amplitude of coupled-channel scattering, $T_{ij}(\sqrt{s})$, taken at a 
total center-of-mass energy $\sqrt{s}$, satisfies the Bethe-Salpeter 
equation~\cite{Kaiser:1995eg,Oset:1998it,Oller:2000fj,Lutz:2001yb}
\begin{align}
    T_{ij}(\sqrt{s})
    =& 
    V_{ij}(\sqrt{s})+V_{il}(\sqrt{s})\,G_{l}(\sqrt{s})\,T_{lj}(\sqrt{s}),
    \label{eq:full}
\end{align}
with the interaction kernel $V_{ij}$ and the meson-baryon loop integral 
$G_i$, and channel indices $i,j$. This set of coupled integral equations 
represents the nonperturbative resummation of $s$-channel loop diagrams. The
solution of Eq.~\eqref{eq:full} is given in matrix form by
\begin{align}
    T
    =& [V^{-1}-G]^{-1} ,
    \nonumber
\end{align}
under the on-shell factorization~\cite{Oset:1998it}.\footnote{One can 
equivalently use the same formulation for the standard integral equation,
regarding the channel indices as intermediate momenta.} This form of the
amplitude is also obtained in the N/D method by neglecting the contributions
from the left-hand cut~\cite{Oller:2000fj}. This guarantees the unitarity of
the scattering amplitude.

In the present work the interaction kernel $V_{ij}$ is identified with the 
leading (Weinberg-Tomozawa) terms derived from the  chiral SU(3) effective 
Lagrangian,
\begin{align}
    V_{ij}(\sqrt{s})
    =&-\frac{C_{ij}}{4f^{2}}
    (2\sqrt{s}-M_i-M_j)\sqrt{\frac{E_i+M_i}{2M_i}}
    \sqrt{\frac{E_j+M_j}{2M_j}} ,
    \label{eq:WTint}  
\end{align}
where $f$ is the pseudoscalar meson decay constant, and $M_i$ and $E_i$ are 
the mass and the energy, respectively, of the baryon in channel $i$. A 
detailed study of interaction terms beyond leading order has been performed 
in Refs.~\cite{Borasoy:2004kk,Borasoy:2005ie}. It was found that such higher
order corrections are relevant for quantitative fine tuning but that the 
essential features of $\bar{K}N$ coupled-channel dynamics can already be 
reproduced at the leading-order level, the strategy that we follow here. 
Effects of higher order terms will be discussed in 
Sec.~\ref{subsec:theoretical}

The coupling strengths $C_{ij}$ in Eq.~\eqref{eq:WTint} are collected in the
matrix
\begin{equation}
    C_{ij}^{I=0}
    =\begin{pmatrix}
       3 & -\sqrt{\frac{3}{2}} & \frac{3}{\sqrt{2}} & 0 \\
         & 4 & 0 & \sqrt{\frac{3}{2}} \\
	 &   & 0 & -\frac{3}{\sqrt{2}} \\
	 &   &   & 3
    \end{pmatrix}
    \nonumber ,
\end{equation}
for the $S=-1$ and $I=0$ channels: $\bar{K}N$ (channel 1), $\pi\Sigma$ 
(channel 2), $\eta\Lambda$ (channel 3), and $K\Xi$ (channel 4). The coupling
strengths in $I=1$ channels are given by
\begin{equation}
    C_{ij}^{I=1}
    =\begin{pmatrix}
       1 & -1 & -\sqrt{\frac{3}{2}} & -\sqrt{\frac{3}{2}} & 0 \\
         & 2 & 0 & 0 & 1 \\
	 &   & 0 & 0 &-\sqrt{\frac{3}{2}} \\
	 &   &   & 0 &-\sqrt{\frac{3}{2}} \\
	 &   &   &   & 1
    \end{pmatrix}
    \nonumber ,
\end{equation}
for the channels $\bar{K}N$ (channel 1), $\pi\Sigma$ (channel 2), 
$\pi\Lambda$ (channel 3), $\eta\Sigma$ (channel 4), and $K\Xi$ (channel 5).

The loop function $G_i(\sqrt{s})$ is given by
\begin{align}
    i\int\frac{d^{4}q}{(2\pi)^{4}}
    \frac{2M_i}{\left[(P-q)^{2}-M_i^{2}+i\epsilon\right]
    \left(q^{2}-m_i^{2}+i\epsilon\right)} 
    \nonumber .
\end{align}
Using dimensional regularization the finite parts of $G_i$ become
\begin{align}
    G_i(\sqrt{s})
    &=\frac{2M_i}{(4\pi)^{2}}
    \Biggl\{a_i(\mu)+\ln\frac{M_i^{2}}{\mu^{2}}
    +\frac{m_i^{2}-M_i^{2}+s}{2s}\ln\frac{m_i^{2}}{M_i^{2}}
    \nonumber\\
    &\quad\quad+\frac{\bar{q}_i}{\sqrt{s}}
    \ln{
    \frac{\phi_{++}(s)\,\phi_{+-}(s)}{\phi_{-+}(s)\,\phi_{--}(s)}
    }
    \Biggr\} ,
    \label{eq:loop}
\end{align}
with
\begin{equation}
\phi_{\pm\pm}(s) = \pm s\pm (M_i^{2}-m_i^{2})+2\bar{q}_i\sqrt{s} ,
    \nonumber\\
\end{equation}
where $a_{i}(\mu)$ are subtraction constants in the channels $i$ and $\mu$ 
is the renormalization scale, $m_i$ is the mass of the meson in channel $i$,
and $\bar{q}_i = \sqrt{[s-(M_i-m_i)^2][s-(M_i+m_i)^2]}/(2\sqrt{s})$ is the 
relevant momentum variable which corresponds to the meson three-momentum in 
the center-of-mass system above threshold.

It has been shown that the scattering amplitude constructed in this way 
reproduces the scattering observables, such as scattering cross sections and
threshold branching ratios. The unitarized amplitude has poles in the 
complex energy plane at the positions of dynamically generated resonances, 
the properties of which are also well described~\cite{Kaiser:1995eg,
Oset:1998it,Oller:2000fj,Lutz:2001yb,Oset:2001cn,Hyodo:2002pk,Hyodo:2003qa,
Borasoy:2004kk,Borasoy:2005ie,Oller:2005ig,Oller:2006jw,Borasoy:2006sr}.

\subsection{Single-channel effective interaction}\label{subsec:single}

\begin{figure*}[tbp]
    \includegraphics[width=\textwidth,clip]{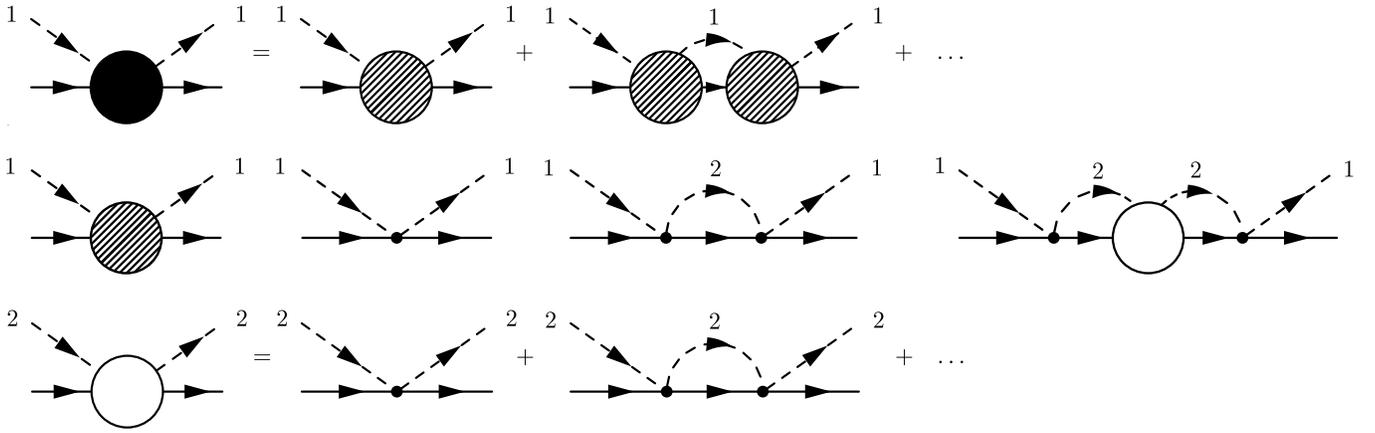} 
    \caption{\label{fig:diagrams2}
    Diagrammatic representation of Eqs.~\eqref{eq:Teffective},
    \eqref{eq:Veffective}, and \eqref{eq:T2single}. The black blob stands 
    for $T^{\text{eff}}=T_{11}$, shaded blobs stand for $V^{\text{eff}}$, 
    and white blobs denote $T^{\text{single}}_{22} $.}
\end{figure*}

In this section we construct an effective interaction in a given single 
channel, the requirement being that the resulting amplitude is identical to 
the solution of the full coupled-channel equations. We start with the 
simplest case of two-channel scattering. The aim is to incorporate the 
dynamics of channel 2 in an effective interaction, $V^{\text{eff}}$, 
operating in channel 1. We would like to obtain the solution $T_{11}$ of 
Eq.~\eqref{eq:full} by solving a single-channel equation with kernel 
interaction $V^{\text{eff}}$, namely,
\begin{align}
    T^{\text{eff}}
    =& \,V^{\text{eff}}+V^{\text{eff}}\,G_1\,T^{\text{eff}}
    \label{eq:Teffective} \\
    =& \,[(V^{\text{eff}})^{-1}-G_1]^{-1} \nonumber \\
    =& \,T_{11} .
    \nonumber 
\end{align}
Consistency with Eq.~\eqref{eq:full} requires that $V^{\text{eff}}$ be the 
sum of the bare interaction, $V_{11}$, in this channel and the contribution 
$\tilde{V}_{11}$ from channel 2:
\begin{align}
    V^{\text{eff}}
    =&\,V_{11} + \tilde{V}_{11}
    \label{eq:Veffective} , \\
    \tilde{V}_{11} 
    =&\,V_{12}\,G_2\,V_{21} +
    V_{12}\,G_2\,T^{\text{single}}_{22}\, G_2
   \, V_{21} \label{eq:Vtilde}\\
    =&\,V_{12}\,G_2\,\left[1+ T^{\text{single}}_{22}\, G_2\right]\,V_{21} 
    \nonumber ,
\end{align}
where $T^{\text{single}}_{22}$ is the single-channel resummation of 
interactions in channel 2:
\begin{align}
    T^{\text{single}}_{22} 
    =&\,V_{22}+V_{22}\,G_2\,T^{\text{single}}_{22} 
    \label{eq:T2single} \\
    =&\,[V_{22}^{-1}-G_2]^{-1}
    \nonumber .
\end{align}
Note that $\tilde{V}_{11}$ includes iterations of one-loop terms in channel 
2 to all orders. If the diagonal component $V_{22}$ is absent, the 
resummation in channel 2 disappears. Therefore, this resummation term 
reflects the effect of the coupled-channel dynamics. 
Equations.~\eqref{eq:Teffective}, \eqref{eq:Veffective}, and 
\eqref{eq:T2single} are diagrammatically illustrated in 
Fig.~\ref{fig:diagrams2}.

\subsection{Multichannel effective interaction}\label{subsec:multi}

It is straightforward to extend this framework to the case with $N$ 
channels. We can generalize Eqs. \eqref{eq:Vtilde} and \eqref{eq:T2single} 
to include the effect of $N-1$ channels $(2,3,\ldots,N)$ into channel 1 as
\begin{align}
    \tilde{V}_{11} 
    =&\sum_{m=2}^{N}V_{1m}\,G_m\,V_{m1}
    +\sum_{m,l=2}^{N}V_{1m}\,G_m\,
    T^{(N-1)}_{ml}\, G_l\,
    V_{l1},
    \label{eq:VtildeNm1} \\
    T^{(N-1)}_{ml}
    =&\, 
    V^{(N-1)}_{ml}+\sum_{k=2}^{N}V^{(N-1)}_{mk}\,G^{(N-1)}_k\,T^{(N-1)}_{kl}
    \nonumber
    \\
    =&\, 
    [(V^{(N-1)})^{-1}-G^{(N-1)}]^{-1} ,~~
    m,l=2,3,\ldots,N.
    \nonumber
\end{align}
The last equation is given as an $(N-1)\times (N-1)$ matrix of the channels
$(2,3,\ldots,N)$. The amplitude of channel 1 is obtained by solving the 
single-channel scattering equation~\eqref{eq:Teffective} with the effective 
interaction of Eqs.~\eqref{eq:Veffective} and \eqref{eq:VtildeNm1}.

In general, we can reduce the $N$-channel problem into effective $n$ 
channels $(1,2,\ldots,n)$ that include the dynamics of $N-n$ channels 
$(n+1,n+2,\ldots,N)$. Starting from the full coupled-channel equation
\begin{equation}
    T_{IJ}^{(N)}
    =V^{(N)}_{IJ}+\sum_{K=1}^{N}V^{(N)}_{IK}\,G^{(N)}_{K}\, T_{KJ}^{(N)}~
    \quad (I,J \in N) ,
    \nonumber
\end{equation}
we want to reproduce the solution of this equation in channels
$(1,2,\ldots,n)$ by the effective interaction $V^{(n)}_{ij}$ as
\begin{align}
    T_{ij}^{(n)}
    =&\,V^{(n)}_{ij}+\sum_{k=1}^{n}V^{(n)}_{ik}\,G^{(n)}_{k}\, T_{kj}^{(n)}
    \quad (i,j \in n)
    \nonumber
    \\
    =&\,\left([(V^{(n)})^{-1}-G^{(n)}]^{-1}\right)_{ij}
    \nonumber \\
    =&\, T_{IJ}^{(N)} \quad( I,J \in n)
    \nonumber .
\end{align}
The effective interaction takes the form
\begin{align}
    V^{(n)}_{ij}
    =&\,V_{ij} + \tilde{V}_{ij}
    \nonumber , \\
    \tilde{V}_{ij} 
    =&\sum_{\alpha=n+1}^{N}\,V_{i\alpha}\,G_{\alpha}\,V_{\alpha j} +
    \sum_{\alpha,\beta =n+1}^{N}V_{i\alpha}\,G_{\alpha} \,
    T^{(N-n)}_{\alpha\beta} \,G_{\beta}\,
    V_{\beta j}
    \nonumber\\
    &\alpha,\beta \in n+1,n+2,\ldots,N ,
    \nonumber 
\end{align}
with $N-n$ channel resummation
\begin{align}
    T^{(N-n)}_{\alpha\beta}
    =&\,V^{(N-n)}_{\alpha\beta}
    +\sum_{\gamma =n+1}^{N}V^{(N-n)}_{\alpha\gamma}\,G^{(N-n)}_{\gamma}\,
    T^{(N-n)}_{\gamma\beta}
    \nonumber \\
    =&\,\left(\left[(V^{(N-n)})^{-1}-G^{(N-n)}
    \right]^{-1}\right)_{\alpha\beta}
    \nonumber , \\
    & \,\alpha,\beta \in n+1,n+2,\ldots,N
    \nonumber 
\end{align}

\section{Effective interaction}\label{sec:effective}

\subsection{Analysis of the $I=0$ $\bar{K}N$ amplitude}
\label{subsec:contribution}

We now turn to our central theme, the construction of the single-channel 
effective $\bar{K}N$ interaction in the isospin basis. There is a strong 
attractive interaction in the $I=0$ channel where the $\Lambda(1405)$ is 
dynamically generated. One expects that a large contribution to the 
$\bar{K}N$ interaction comes from the $\pi\Sigma$ channel. It is therefore 
useful and instructive to compare the effective interaction in the 
$\bar{K}N$-$\pi\Sigma$ coupled-channel case (two-channel model) with that 
including four coupled channels (full model).

The parameters in Eqs.~\eqref{eq:WTint} and \eqref{eq:loop} are fixed as 
$f=106.95 $ MeV, $\mu=630$ MeV, and $a_i=-1.96$ for all channels. We use the
physical hadron masses averaged over isospin multiplets. As shown in 
Refs.~\cite{Hyodo:2002pk,Hyodo:2003qa}, the model with these parameters 
reproduces the experimental observables such as total cross sections for 
elastic and inelastic $K^-p$ scattering, threshold branching ratios, and the
$\pi\Sigma$ mass spectrum in the region of the $\Lambda(1405)$. 

In what follows we present forward scattering amplitudes in units of 
femtometers, related to the amplitudes $T$ in Eq.~\eqref{eq:full} by
\begin{equation}
    F_{\bar{K}N} = -\frac{M_N}{4\pi \sqrt{s}}\, T_{11} ,~~ 
    F_{\pi\Sigma} = -\frac{M_\Sigma}{4\pi \sqrt{s}}\, T_{22},
    \nonumber
\end{equation}
etc. With this commonly used convention, scattering lengths are directly 
given by the values of $F_i$ at threshold.

\begin{figure*}[tbp]
    \centering
    \includegraphics[width=0.75\textwidth,clip]{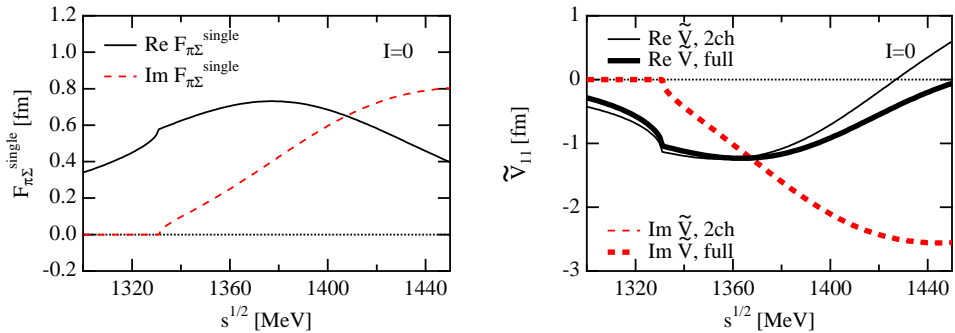}
    \caption{\label{fig:contribution1}
    (Color online) The single-channel $\pi\Sigma$ scattering amplitude
    $F_{\pi\Sigma}^{\text{single}} 
    = -(4\pi M_\Sigma/\sqrt{s})\,T_{22}^{\text{single}}$ (left panel) and 
    the interaction term $\tilde{V}_{11}$ (right panel). Real parts are 
    shown as solid lines; imaginary parts are represented as dashed lines. 
    The thin lines are the result of the two-channel model and the thick 
    lines are the result of the full (four-channel) model. The imaginary 
    parts of $\tilde{V}_{11}$ in the two-channel and full models are 
    indistinguishable in this figure.}
\end{figure*}

Let us examine separately the contributions to the effective interaction 
$V^{\text{eff}}$. The left panel of Fig.~\ref{fig:contribution1} shows the 
real and imaginary parts of the amplitude with $\pi\Sigma$ single-channel 
resummation [$T_{22}^{\text{single}}$ in Eq.~\eqref{eq:T2single}]. At first 
sight there appears to be no prominent resonance structure in this 
amplitude, but it nevertheless develops a pole in the complex energy plane 
at
\begin{equation}
    z_2(\pi\Sigma \text{ only}) = 1388 - 96 i \text{ MeV} .
    \label{eq:z2single}
\end{equation}
The large imaginary part, in spite of the relatively small phase space 
($\sim$50 MeV above the threshold) is a special feature of this $s$-wave 
$\pi\Sigma$ single-channel resonance.

In the right panel of Fig.~\ref{fig:contribution1}, the $\bar{K}N$ 
interaction $\tilde{V}_{11}$ from coupled-channel dynamics 
[Eqs.~\eqref{eq:Vtilde} and \eqref{eq:VtildeNm1}] are plotted. Thin lines 
represent the results with the two-channel model; the results of the full 
model are also shown as thick lines. 

One also observes that the imaginary part of $\tilde{V}_{11}$ is almost 
identical in the two-channel and full models. The reason is as follows. 
First, the imaginary part of $\tilde{V}_{11}$ comes only from the loop of 
the $\pi\Sigma$ channel $G_2$ in this energy region, when we expand the 
amplitude. Taking into account the zeros in the coupling strengths 
$C_{23}=C_{32}=C_{14}=C_{41}=0$, one can show that the difference between 
$\im \tilde{V}_{11}^{(2)}$ and $\im \tilde{V}_{11}^{(4)} $ is of the order 
of $\mathcal{O}[V(GV)^3]$. However, the magnitude of $GV$ is roughly 
estimated as $\mathcal{O}(10^{-1})$. Numerically, the magnitudes of the real
and imaginary parts are smaller than 0.7 for the relevant energy region. 
Therefore, terms with $\mathcal{O}[V(GV)^3]$ are much smaller in magnitude 
than the imaginary part of $\tilde{V}_{11}$, which is of order 
$\mathcal{O}[V(GV)]$. Hence the difference between 
$\im \tilde{V}_{11}^{(4)}$ and $\im \tilde{V}_{11}^{(2)}$ is indeed small.

\begin{figure*}[tbp]
    \centering
    \includegraphics[width=0.75\textwidth,clip]{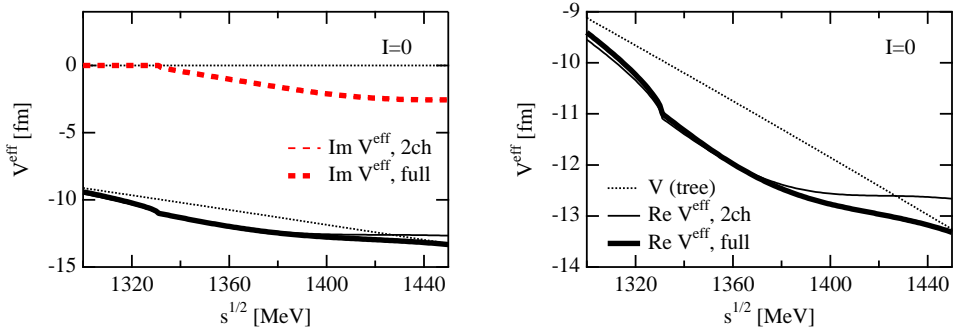}
    \caption{\label{fig:contribution2}
    (Color online) The $\bar{K}N$ $(I=0)$ interaction at tree level given by
    the Weinberg-Tomozawa term (dotted lines), the effective interaction in 
    the two-channel model (thin lines), and the effective interaction in the
    full model (thick lines). The real parts are shown by the solid lines, 
    and the imaginary parts are depicted by the dashed lines. The right 
    panel details the upper part of the left panel. The imaginary parts of 
    $V^{\text{eff}}$ in the two-channel and full models are 
    indistinguishable in this figure.}
\end{figure*}

The effective $\bar{K}N$ interaction $V^{\text{eff}}$ is plotted in 
Fig.~\ref{fig:contribution2}, together with the tree-level Weinberg-Tomozawa
term for the $\bar{K}N$ channel. By construction of the effective 
interaction [Eq.~\eqref{eq:Veffective}], the difference from the tree-level 
one is
attributed to the coupled-channel dynamics. As seen in the figure, the 
$\pi\Sigma$ and other coupled channels enhance the strength of the 
interaction at low energy, although not by a large amount. The primary 
difference is seen in the energy dependence of the interaction kernel. The 
imaginary part is smaller in magnitude than the real part. This permits 
treating the imaginary part perturbatively in the effective $\bar{K}N$ 
interaction.

\begin{figure*}[tbp]
    \centering
    \includegraphics[width=0.75\textwidth,clip]{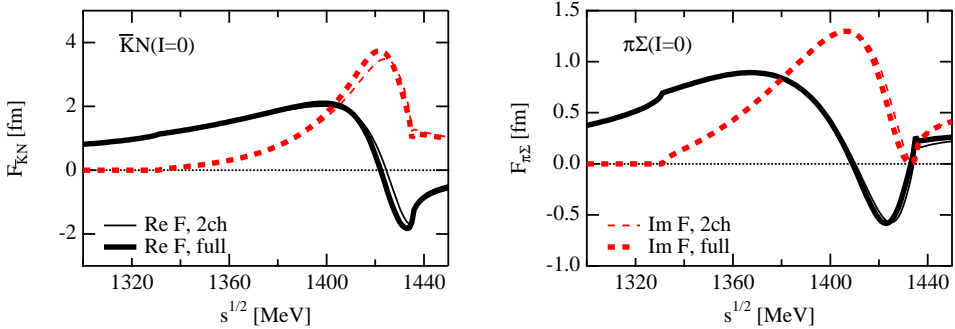}
    \caption{\label{fig:Full}
    (Color online) Scattering amplitudes $F_{\bar{K}N}$ channel (left) and 
    $F_{\pi\Sigma}$  (right) in the $(I = 0)$ two-channel model (thin lines)
    and in the full model (thick lines). Real parts are shown as solid 
    lines and imaginary parts as dashed lines.
    }
\end{figure*}

In the left panel of Fig.~\ref{fig:Full}, we show the result of $\bar{K}N$ 
scattering amplitude $T^{\text{eff}}$, obtained by solving the 
single-channel scattering equation with $V^{\text{eff}}$. The full amplitude
in the $\pi\Sigma$ channel is plotted in the right panel for comparison. We 
numerically checked that $T^{\text{eff}}$ coincides with $T_{11}$ resulting 
from the coupled-channel equations. 

In the $\bar{K}N$ scattering amplitude, the resonance structure is observed 
at around 1420 MeV, significantly higher than the nominal position of the 
$\Lambda(1405)$. However, the peak in the $\pi\Sigma$ amplitude (shown in 
the right panel of Fig.~\ref{fig:Full}) is in fact located close to 1405 
MeV. This is a consequence of the two-pole structure~\cite{Jido:2003cb,
Hyodo:2003jw,Magas:2005vu}, which we will discuss in detail in the following
sections. Results with the full model (thick lines) are not very much 
different from those with the two-channel model (thin lines). This indicates
that the scattering amplitude around the $\bar{K}N$ threshold is well 
described by the $\bar{K}N$-$\pi\Sigma$ coupled-channel system, and we 
confirm that the $\eta \Lambda$ and $K\Xi$ channels are unimportant for the 
physics of the $\bar{K}N$ interaction in the energy region of interest.

\subsection{Structure of $\Lambda(1405)$}
\label{subsec:L1405}

We now discuss the pole structure of the $\Lambda(1405)$ resonance in 
greater detail. The $\bar{K}N$ scattering amplitude $T^{\text{eff}}
(\sqrt{s})$, obtained by using the two-channel model, develops two poles at
\begin{align*}
    z_1^{(2)} &= 1432 - 17 i \text{ MeV}, \quad
    z_2^{(2)} = 1398 - 73 i \text{ MeV} .
\end{align*}
These pole positions move slightly, to
\begin{align*}
    z_1^{(4)} &= 1428 - 17 i \text{ MeV}, \quad
    z_2^{(4)} = 1400 - 76 i \text{ MeV} ,
\end{align*}
in the full model with four channels. Again, the deviation from the 
two-channel model is only marginal. We thus confirm the dominance of the 
$\bar{K}N$-$\pi\Sigma$ coupled-channel dynamics in the $\bar{K}N$ amplitude.
As discussed in Ref.~\cite{Jido:2003cb}, the poles $z_1$ and $z_2$ have 
different coupling strengths to the $\pi\Sigma$ and $\bar{K}N$ channels, 
leading to the different shapes in the $\bar{K}N$ and $\pi\Sigma$ 
amplitudes, as seen in Fig.~\ref{fig:Full}.

\begin{figure}[tbp]
    \centering
    \includegraphics[width=0.5\textwidth,clip]{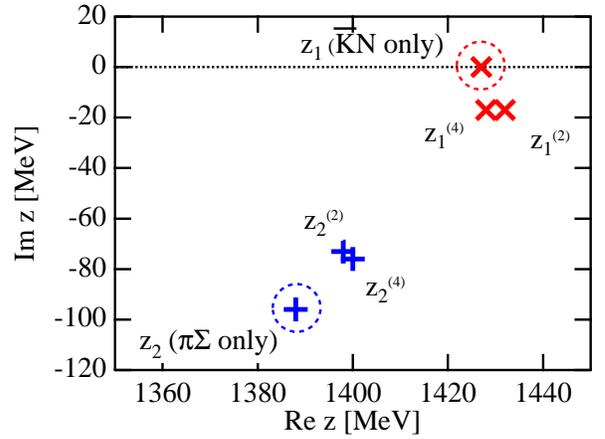}
    \caption{\label{fig:pole}
    (Color online) Pole positions of the $\bar{K}N(I=0)$ scattering 
    amplitude resulting from the single-channel, two-channel, and full 
    (four-channel) models.}
\end{figure}

Next we study the origin of these poles. As previously mentioned, there is a
pole [Eq.~\eqref{eq:z2single}] in the amplitude $T_{22}$ representing 
$\pi\Sigma$ single-channel resummation. We can also perform the resummation 
of the tree-level Weinberg-Tomozawa term in the single $\bar{K}N$ channel, 
which generates a bound state pole at
\begin{equation}
    z_1(\bar{K}N \text{ only}) = 1427  \text{ MeV} .
    \nonumber
\end{equation}
The pole positions for single-channel, two-channel, and full models are 
plotted in Fig.~\ref{fig:pole}. These positions obviously suggest that the
pole $z_1(\bar{K}N \text{ only})$ is the origin of the poles $z_1^{(2)}$ and
$z_1^{(4)}$, whereas $z_2(\pi\Sigma \text{ only})$ is the origin of the 
poles $z_2^{(2)}$ and $z_2^{(4)}$. This observation agrees once again with 
the qualitative behavior discussed in Refs.~\cite{Jido:2003cb,
Garcia-Recio:2003ks} that the pole $z_1$ strongly couples to the $\bar{K}N$ 
channel and the pole $z_2$ to the $\pi\Sigma$ channel.

The principal features from chiral SU(3) dynamics underlying this behavior 
are as follows. The driving (attractive) $s$-wave interactions in the 
$\bar{K}N$ and $\pi\Sigma$ channels are determined by the Goldstone boson 
nature of the pseudoscalar octet mesons. In the chiral limit (i.e., with all
quark masses $m_{u,d}$ and $m_s$ strictly equal to zero) these interactions 
would all vanish at threshold, with massless kaon and pion. Chiral symmetry 
dictates that their interaction strength grows linearly (in leading order) 
with their energy. Explicit chiral symmetry breaking  gives those mesons 
their masses and moves the meson-baryon threshold energies to their physical
values. At the kaon-nucleon threshold, the leading $\bar{K}N$ interaction, 
$V_{11} \sim -\frac{3m_K}{2f^2}$, becomes sufficiently attractive to produce
a weakly bound state. At the $\pi\Sigma$ threshold the corresponding leading
$\pi\Sigma$ interaction, $V_{22} \sim -\frac{2m_\pi}{f^2}$, is too weak to 
support a $\pi\Sigma$ bound state, but it still generates a resonance above 
threshold. 

At a more quantitative level, the driving attractive (Weinberg-Tomozawa) 
interaction terms in the  $S = -1$ meson-baryon channels, with their 
characteristic energy dependence,
\begin{equation}
   V_{11} \equiv V_{\bar{K}N}
   \simeq -\frac{3}{2f^2}\left(\sqrt{s} - M_N\right)
    \nonumber
\end{equation}
and
\begin{equation}
   V_{22} \equiv V_{\pi\Sigma}
   \simeq -\frac{2}{f^2}\left(\sqrt{s} - M_\Sigma\right),
    \nonumber
\end{equation}
generate a $\bar{K}N$ bound state and a $\pi\Sigma$ resonance already in the
absence of channel couplings (i.e., for $V_{12} = V_{21} = 0$). In the 
typical energy range of $\sqrt{s} \sim 1410$--$1420$ MeV of interest here, 
the ratio of these driving interactions,
\begin{equation}
   \frac{V_{\bar{K}N}}{V_{\pi\Sigma}}
   \sim -\frac{3}{4}\left(\frac{\sqrt{s} - M_N}{ \sqrt{s} - M_\Sigma}\right)
    \nonumber
\end{equation}
is about 1.6. The $\bar{K}N$ interaction is effectively stronger, but the 
$\pi\Sigma$ interaction is sizable and cannot be ignored even though the 
$\pi\Sigma$ resonance pole is located at a considerable distance from the 
real axis in the complex energy plane. The isolated $\bar{K}N$ bound state, 
in contrast, has a binding energy of only 5 MeV at this stage. By turning
on the nondiagonal $\bar{K}N \leftrightarrow\pi\Sigma$ couplings, 
$V_{12} = V_{21}$, both the $\pi\Sigma$ resonance and $\bar{K}N$ bound state
poles move to their final positions as shown in Fig.~\ref{fig:pole}. The 
$\bar{K}N$ bound state turns into a quasibound state embedded in the 
$\pi\Sigma$ continuum, with a decay width of about 35 MeV. At the same time,
by its coupling to the (virtual) $\bar{K}N$ channel, the previously isolated
$\pi\Sigma$ resonance shifts and reduces its width by about $20\%$, from 
$\Gamma_{\pi\Sigma}\simeq 190$ MeV to $\Gamma_{\pi\Sigma}\simeq 150$ MeV. 

These considerations imply a subtlety in assigning a mass (or a $\bar{K}N$ 
binding energy) to the $\Lambda(1405)$. Empirically, the only information at
hand is the $\pi\Sigma$ mass spectrum given by the imaginary part of 
$T_{22} \equiv T_{\pi\Sigma}$ [see Fig.~\ref{fig:Full} (right)]. This mass 
spectrum has its maximum indeed at $\sqrt{s} \simeq 1405$ MeV. Although its 
spectral shape is far from that of a Breit-Wigner resonance form, one may 
nevertheless read off a width of about 50 MeV. This is the mass and width 
assignment given to the $\Lambda(1405)$ in the Particle Data Group table. 
However, the amplitude $T^{\text{eff}} = T_{11} \equiv T_{\bar{K}N}$ in the 
$\bar{K}N$ channel shown in Fig.~\ref{fig:Full} (left) in the form of 
$F_{\bar{K}N}$ has evidently quite different features. This is the amplitude
relevant for subthreshold extrapolations of the $\bar{K}N$ interaction. The 
$\bar{K}N$ quasibound state, signaled by the zero of Re $F_{\bar{K}N}$, is 
seen to be located at $\sqrt{s}\simeq 1420$ MeV, {\it not} at 1405 MeV, and 
almost coincides with the maximum of Im $F_{\bar{K}N}$. The actual 
$\bar{K}N \rightarrow\pi\Sigma$ decay width is about 20$\%$ smaller than the
one naively identified with the breadth of the $\pi\Sigma$ mass spectrum. 
One must therefore conclude that the $\bar{K}N$ quasibound state, commonly 
associated with the $\Lambda(1405)$, has a binding energy of only about 12 
MeV (and {\it not} the 27 MeV often used to tune phenomenological $\bar{K}N$
potentials).\footnote{In our present calculations, the isospin 
averaged $\bar{K}$ mass has been used in practice, so that the 
$\bar{K}N$ threshold actually appears at $\sqrt{s}=1435$ MeV, shifted 
by 3 MeV from the $K^-p$ threshold. The value of the binding energies 
just mentioned are thus understood to be shifted by the same amount.}

Given the obvious relevance of this discussion to the existence (or 
nonexistence) of deeply bound $\bar{K}$-nuclear states, it is now important
to estimate theoretical uncertainties and examine possible ambiguities.

\subsection{Theoretical uncertainties and $I=1$ amplitude}
\label{subsec:theoretical}

Detailed investigations have been performed concerning the position of the 
second pole $z_2$, especially its sensitivity to higher order terms in the 
chiral effective Lagrangian \cite{Borasoy:2004kk,Borasoy:2005ie,
Oller:2005ig,Oller:2006jw,Borasoy:2006sr}. This section presents a 
conservative assessment of such uncertainties, examining different chiral 
coupled-channel calculations in comparison.

Several variants of the chiral unitary approach will be used in this test,
all of which start from a leading Weinberg-Tomozawa (WT) term in the 
interaction kernel but differ in their detailed treatment of subtraction 
constants. The differences among these models have their origin in the 
fitting procedures to experimental data, primarily through ambiguities of 
the $\pi\Sigma$ mass spectrum with its limited data quality.

Oset-Ramos-Bennhold (ORB)~\cite{Oset:2001cn} determine the subtraction
constants by matching the loop function with that obtained in the 
three-momentum cutoff, with which they successfully reproduce the 
observables~\cite{Oset:1998it}. Hyodo-Nam-Jido-Hosaka 
(HNJH)~\cite{Hyodo:2002pk} used one single subtraction constant in all 
channels to fit the data. A systematic $\chi^2$ fit was performed by 
Borasoy-Ni\ss ler-Weise (BNW)~\cite{Borasoy:2005ie} and by 
Borasoy-Mei\ss ner-Ni\ss ler~\cite{Borasoy:2006sr}, where all the 
subtraction constants were used to fit the experimental data and the 
influence of higher order terms in the chiral effective Lagrangian has been 
studied. The fitted observables in these investigations are total cross 
sections of $K^-p$ scattering in elastic and inelastic channels, threshold 
branching ratios, invariant mass distribution in the $\pi\Sigma$ channel and
the $K^- p$ scattering length deduced from kaonic hydrogen data. The 
analysis in the preceding sections is based on the simpler HNJH model.

To estimate systematic theoretical uncertainties, we adopt all these models 
and derive the corresponding effective interactions. The definition of the 
subtraction constant $a^{\text{Borasoy}}$ in Refs.~\cite{Borasoy:2005ie,
Borasoy:2006sr} is related to the present convention (and those in 
Refs.~\cite{Oset:2001cn,Hyodo:2002pk}) by
\begin{equation}
    a(\mu) = 16\pi^2a^{\text{Borasoy}}(\mu)-1 .
    \nonumber
\end{equation}
Changes of the renormalization scale in different models are related by
\begin{equation}
    a(\mu^{\prime})
    =a(\mu)+2\ln (\mu^{\prime}/\mu) ,
    \nonumber
\end{equation}
where $\mu=630$ MeV. The subtraction constants determined in the models 
under consideration are shown in Table~\ref{tbl:parameters} together with 
the corresponding values for the meson decay constant $f$.\footnote{For the 
ORB model, it was stated that $f=1.15\times f_{\pi}$ in 
Ref.~\cite{Oset:2001cn}, but in the actual calculation 
$f=1.123\times f_{\pi}$ was used, as noted in Ref.~\cite{Jido:2003cb}. For 
the BNW model, there is a misprint in the column ``WT term'' in Table 1 of 
Ref.~\cite{Borasoy:2005ie}: The signs of all subtraction constants should be
inverted.} We use isospin-averaged hadron masses as input. One should note 
that the results presented here are not exactly identical to those in the 
original papers because of differences in the input masses, isospin breaking
effects, and so on. The accuracy of the analysis is nonetheless sufficient 
for the present purpose of estimating theoretical uncertainties.

\begin{table*}[tbp]
    \centering
    \caption{Subtraction constants $a_i$ at $\mu=630$ MeV and meson decay 
    constant $f$ in different models.}
    \begin{ruledtabular}
    \begin{tabular}{llrrrrrr}
        Reference & $f$ (MeV) & $a_{\bar{K}N}$ & $a_{\pi\Sigma}$ 
	& $a_{\eta\Lambda}$
	& $a_{K\Xi}$ & $a_{\pi\Lambda}$ & $a_{\eta\Sigma}$ \\
        \hline
        ORB~\cite{Oset:2001cn} & 103.7652 & $-1.84$ & $-2.00$
	 & $-2.25$ & $-2.67$ & $-1.83$ & $-2.38$  \\
        HNJH~\cite{Hyodo:2002pk} & 106.95 & $-1.96$ & $-1.96$
	 & $-1.96$ & $-1.96$ & $-1.96$ & $-1.96$ \\
        BNW~\cite{Borasoy:2005ie} & 111.2& $-1.86$ & $-2.35$
	 & $-2.67$ & $-2.62$ & $-1.23$ & $-2.80$  \\
        BMN~\cite{Borasoy:2006sr} & 120.9 & $-2.21$ & $-2.38$
	 & $-2.19$ & $9.59$ & $-3.88$ & $-2.15$  \\
    \end{tabular}
    \end{ruledtabular}
    \label{tbl:parameters}
\end{table*}

\begin{figure*}[tbp]
    \centering
    \includegraphics[width=0.75\textwidth,clip]{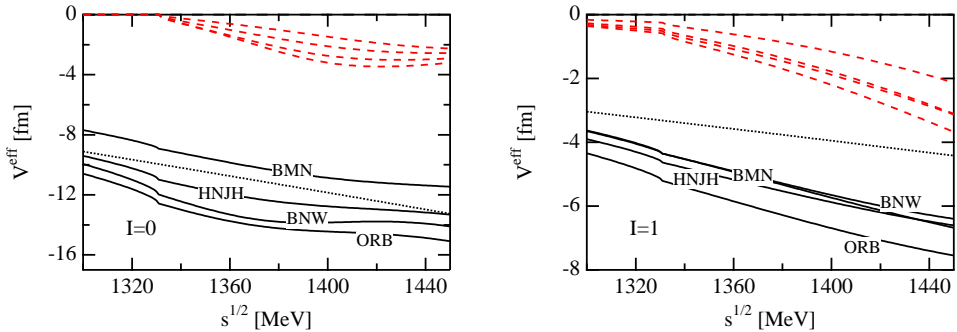}
    \caption{\label{fig:Effective_model}
    (Color online) The effective interaction $V^{\text{eff}}$ of the 
    $\bar{K}N(I=0)$ channel (left) and that of the $\bar{K}N(I=1)$ channel 
    (right) in different models. The real parts are shown as solid lines 
    and imaginary parts as dashed lines. The lines correspond to the models 
    as indicated in the figure. The dotted line is the tree-level WT 
    interaction.}
\end{figure*}

The effective interactions calculated with these models are shown in 
Fig.~\ref{fig:Effective_model} together with the tree-level WT term result 
(with $f\simeq107$ MeV, a value intermediate between the empirical pion and 
kaon decay constants). The strengths of the effective interactions are 
roughly comparable with the WT term for the $\bar{K}N(I=0)$ channel, whereas
the coupled-channel dynamics in the $\bar{K}N(I=1)$ channel enhances the 
interaction strengths by about 50\% from that of the WT term.

\begin{figure*}[tbp]
    \centering
    \includegraphics[width=0.75\textwidth,clip]{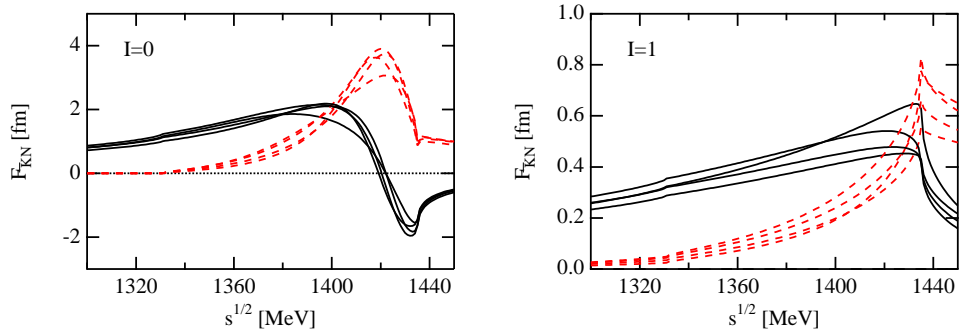}
    \caption{\label{fig:Full_model}
    (Color online) Scattering amplitude $F_{\bar{K}N} = 
    -(4\pi M_N/\sqrt{s})T^{\text{eff}}$ of the $\bar{K}N(I=0)$ channel 
    (left) and of the $\bar{K}N(I=1)$ channel (right) in different models. 
    The real parts are shown as solid lines and imaginary parts as dashed 
    lines.}
\end{figure*}

With these effective interactions we find the scattering amplitudes shown in
Fig.~\ref{fig:Full_model}. The model dependence of these amplitudes is not 
large despite the differences among the underlying effective interactions.
This is understandable because these interactions are all fitted to similar 
data sets. The differences in the interaction strengths among different 
models are in large part compensated by corresponding differences in the 
subtraction constants. Note that the peak position of the imaginary part of 
the $\bar{K}N(I=0)$ amplitude is around 1420 MeV in all models. This 
observation is also consistent with the solution of the Lippmann-Schwinger 
equation obtained by using a chiral interaction approximated by a separable 
potential~\cite{Kaiser:1995eg,Ikeda:2007nz}. 

\begin{figure}[tbp]
    \centering
    \includegraphics[width=0.5\textwidth,clip]{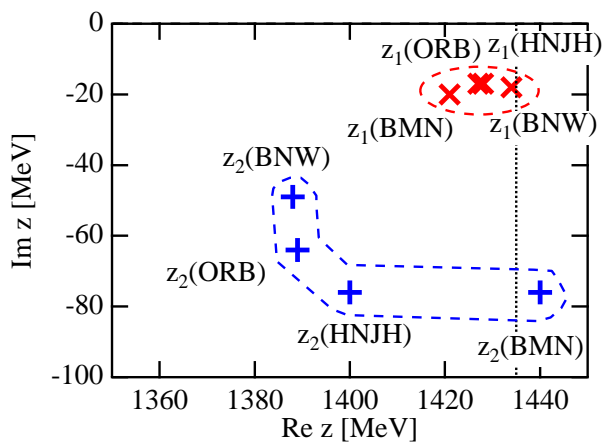}
    \caption{\label{fig:pole_model}
    (Color online) Pole positions of the $\bar{K}N(I=0)$ scattering 
    amplitude in different models. The dotted line denotes the threshold of 
    the $\bar{K}N$ channel.}
\end{figure}

Calculated $K^-p$ scattering lengths $a_{K^-p}=(a_{\bar{K}N(I=0)}+
a_{\bar{K}N(I=1)})/2$ and pole positions of the scattering amplitudes are 
summarized in Table~\ref{tbl:lengths}. The pole positions are also plotted 
in Fig.~\ref{fig:pole_model}. The pole $z_2$ in BMN~\cite{Borasoy:2006sr} is
found above the $\bar{K}N$ threshold. It is located on the Riemann sheet, 
which is unphysical for $\pi\Sigma$ and physical for $\bar{K}N$. For 
$\im z<0$ this sort of pole does not directly influence the physical 
scattering amplitudes.

\begin{table}[tbp]
    \centering
    \caption{$K^-p$ scattering lengths and pole positions in different 
    models.}
    \begin{ruledtabular}
    \begin{tabular}{lccc}
        Reference & $a_{K^-p}$ (fm) & $z_1$ (MeV) & $z_2$ (MeV)  \\
        \hline
        ORB~\cite{Oset:2001cn} & $-0.617 + 0.861 i$
	& $1427-17i$ & $1389-64i$ \\
        HNJH~\cite{Hyodo:2002pk} & $-0.608 + 0.835 i$ & $1428 -17i$ & 
	$1400-76i$ \\
        BNW~\cite{Borasoy:2005ie} & $-0.532 + 0.833 i$
	& $1434-18i$ & $1388-49i$   \\
        BMN~\cite{Borasoy:2006sr} & $-0.410 + 0.824 i$
	& $1421-20i$ & $1440-76i$   \\
    \end{tabular}
    \end{ruledtabular}
    \label{tbl:lengths}
\end{table}

The pole positions of $z_2$ are scattered over a wide range of the complex 
energy plane depending on the model used. Moreover, the detailed behavior of
this pole is sensitive to physics beyond leading (WT) order in the chiral 
effective Lagrangian, studied systematically in Ref.~\cite{Borasoy:2005ie}. 
However, such higher order corrections can partially be absorbed by 
readjusting the pseudoscalar decay constant $f$.       

In contrast to the strong model dependence of $z_2$ the location of the pole
$z_1$ is quite stable, with Re $z_1$ positioned in a narrow window around 
1420--1430 MeV. In spite of the differences in the position of the pole 
$z_2$, the $\bar{K}N(I=0)$ amplitudes in Fig.~\ref{fig:Full_model} do not 
change very much. This is because the behavior of the $\bar{K}N$ amplitude 
is largely determined by the contribution from the pole $z_1$.

\begin{figure}[tbp]
    \centering
    \includegraphics[width=0.5\textwidth,clip]{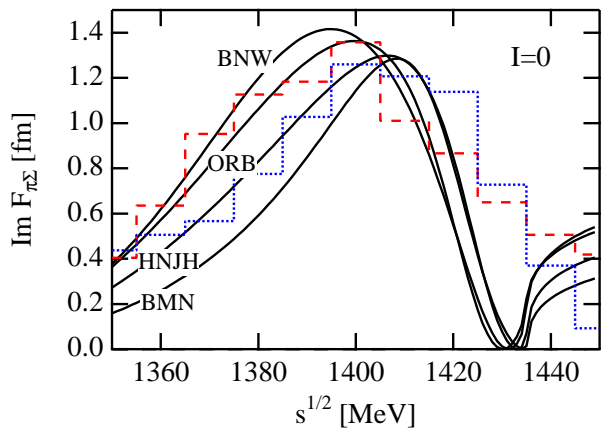}
    \caption{\label{fig:pSdist}
    (Color online) Imaginary part of the $\pi\Sigma(I=0)$ amplitude in 
    different models. The dashed histogram indicates data from 
    Ref.~\cite{Thomas:1973uh} ($\pi^-\Sigma^+$ spectrum) and the dotted 
    histogram indicates data from Ref.~\cite{Hemingway:1985pz} (sum of 
    $\pi^{\pm}\Sigma^{\mp}$ spectra).}
\end{figure}

The spread in the $z_2$ pole positions can be observed in the $\pi\Sigma$ 
amplitude. In Fig.~\ref{fig:pSdist} we plot the imaginary part of the 
$\pi\Sigma (I=0)$ amplitude where the maximum of the spectrum is commonly
identified with the  $\Lambda(1405)$ resonance. The model dependence of the 
$\pi\Sigma$ amplitudes is stronger than that of the $\bar{K}N$ amplitude, 
reflecting the contribution from the pole $z_2$. For comparison, we plot the
invariant mass spectrum of $\pi^-\Sigma^+$ in Ref.~\cite{Hemingway:1985pz} 
and the sum of $\pi^{\pm}\Sigma^{\mp}$ in Ref.~\cite{Thomas:1973uh}. 

One should be careful when comparing the calculated $I = 0$ $\pi\Sigma$ mass
spectrum with experimental data. In principle we need all three $\pi\Sigma$ 
states ($\pi^{\pm}\Sigma^{\mp}$, $\pi^0\Sigma^0$) \textit{simultaneously} to
construct the pure $I=0$ spectrum, since there are three isospin 
states: $I=0$, 1, and 2.\footnote{In practice, $\pi^0\Sigma^0$ has no $I=1$ 
component and the contribution from $I=2$ is considered to be small.} The 
available experimental $\pi\Sigma$ spectra were generated in one or two 
final states for each experiment (sum of charged 
states~\cite{Thomas:1973uh}, 
$\pi^-\Sigma^+$~\cite{Hemingway:1985pz}, 
and charged states~\cite{Ahn:2003mv},
$\pi^0\Sigma^0$~\cite{Prakhov:2004an}, 
$\pi^0\Sigma^0$~\cite{Zychor:2007gf}).
This means in fact that the $\pi\Sigma$ mass spectrum representing the $I=0$
channel has not been extracted so far. The value quoted by the Particle Data
Group (PDG)~\cite{\PDG} is based only on the analysis of 
Ref.~\cite{Dalitz:1991sq} in which the $\pi^-\Sigma^+$ spectrum of 
Ref.~\cite{Hemingway:1985pz} was fitted by the $I=0$ amplitudes of 
theoretical models. Even if the $I=1$ and $I=2$ components are smaller than 
the $I=0$ spectrum, cross terms such as Re$T^0T^{1*}$ may distort the shape 
of the spectrum~\cite{Nacher:1998mi,Hyodo:2004vt}.

Additional theoretical uncertainties concern the chiral coupled-channel 
approach itself. The present investigation is focused on chiral SU(3) models
with leading-order interaction. Effects of higher order $\bar{K}N$ couplings
have been studied systematically in Ref.~\cite{Borasoy:2005ie}. 
Representative examples are listed in Table~\ref{tbl:estimate}, which shows 
the values of Re $F_{\bar{K}N}(I=0)$ at a given subthreshold energy,
$\sqrt{s} = 1360$ MeV. The case WT is equivalent to the leading-order 
results discussed in this paper. Cases c and s include next-to-leading-order
terms in the chiral SU(3) interaction kernel. In all cases the input 
parameters are adjusted to reproduce the available experimental data. The 
results with inclusion of higher order terms tend to increase slightly the 
strength of the $\bar{K}N$ amplitude in the far-subthreshold region, within 
a limited uncertainty bound of about 20 \%.

\begin{table}[tbp]
    \centering
    \caption{The real parts of the amplitude $F_{\bar{K}N}(I=0)$ at 
    $\sqrt{s}=1360$ MeV in comparison with the next-to-leading-order 
    results.}
    \begin{ruledtabular}
    \begin{tabular}{lcc}
        Model & Re $F_{\bar{K}N}(I=0)$ (fm) & Order  \\
        \hline
        this work & $1.6\pm 0.2$ & leading   \\
        Ref.~\cite{Borasoy:2005ie} WT & $\sim$$1.6$ & leading  \\
        Ref.~\cite{Borasoy:2005ie} c & $\sim$$1.9$ & $p^2$  \\
        Ref.~\cite{Borasoy:2005ie} s & $\sim$$2.0$ & $p^2$  \\
    \end{tabular}
    \end{ruledtabular}
    \label{tbl:estimate}
\end{table}

Early coupled-channel calculations~\cite{Kaiser:1995eg,Kaiser:1997js} 
suggested larger subthreshold values of Re $F_{\bar{K}N}$ (e.g., 
$\sim$$3.8$ fm~\cite{Kaiser:1995eg} and $\sim$$3.0$ fm~\cite{Kaiser:1997js} 
at $\sqrt{s}=1360$ MeV). These differences in comparison with the more 
recent chiral coupled-channel results can presumably be traced to the use of
the (nonrelativistic) Lippmann-Schwinger equation in combination with 
separable approximations and corresponding cutoffs for the interaction 
kernels in the early approaches, as opposed to the Bethe-Salpeter equation 
and dispersion relation techniques applied in the more recent computations. 
A common feature of all approaches is the location of the zero of Re 
$F_{\bar{K}N}$ \textit{above} the canonical 1405 MeV, irrespective of the 
detailed extrapolation of the amplitude to the far-subthreshold region. 

Further improvements should incorporate rigorous theoretical constraints on 
the subthreshold $\bar{K}N$ amplitude, such as crossing 
symmetry~\cite{Martin:1980qe}, as discussed within the framework of a chiral
coupled-channel approach in Ref.~\cite{Lutz:2001yb}. In addition, a detailed
consistency analysis of subtraction constants in view of constraints from 
order $p^3$ counter terms, as elaborated in Ref.~\cite{Lutz:2001yb}, is 
certainly desirable along the same line as the present investigation.

Although the primary uncertainty of any detailed $\bar{K}N$ subthreshold 
extrapolation is evidently rooted in the lack of sufficiently accurate data 
for the $\pi\Sigma$ spectral functions, it is nevertheless remarkable that 
most chiral models consistently agree within limited errors on the shapes 
and magnitudes of real and imaginary parts of the $\bar{K}N$ amplitudes, 
in both $I = 0$ and $I = 1$ channels. The following intermediate conclusions
can therefore be drawn:
\begin{itemize}
    \item{The position of the $\Lambda(1405)$ as an $I = 0$ quasibound 
    $\bar{K}N$ state embedded in the $\pi\Sigma$ continuum, when identified 
    with the zero of Re $T^{\text{eff}}(\bar{K}N)$, is located at 
    $\sqrt{s} \simeq 1420$ MeV, {\it not} at 1405 MeV. The corresponding 
    $K^- p$ ``binding energy" is thus 12 MeV instead of 27 MeV.}
    \item{ The $\pi\Sigma$ mass spectrum (with so far very limited accuracy 
    of the existing data base) reflects primarily coupled-channel dynamics 
    around the $z_2$ pole in the $\pi\Sigma$ amplitude. The maximum of the 
    $\pi\Sigma$ mass spectrum around $\sqrt{s} \sim (1400$-$1410)$ MeV is 
    therefore {\it not} to be directly interpreted as the position of the 
    $\Lambda(1405)$.}
\end{itemize}

\section{``Equivalent'' local potential}\label{sec:local}

\subsection{Local pseudopotential from the \\effective $\bar{K}N$ 
interaction}\label{subsec:formulation}

Next we construct an equivalent local $\bar{K}N$ pseudopotential in 
coordinate space. Such a potential is, for example, a useful input to 
computations of $\bar{K}$-nuclear few-body systems. Equivalence means that 
the solution of the Schr\"odinger or Lippmann-Schwinger equation with this 
pseudopotential should approximate the scattering amplitude derived from the
full chiral coupled-channel calculation as closely as possible.

Consider an $s$-wave antikaon-nucleon system in nonrelativistic quantum 
mechanics. The Schr\"odinger equation for the radial wave function $u(r)$ is
(with $\hbar$ set to 1)
\begin{align}
    -\frac{1}{2\mu}\frac{d^2 u(r)}{d r^2}
    +U(r,E)\, u(r)
    =&E\, u(r) .
    \label{eq:schroedinger}
\end{align}
The potential $U(r,E)$ is complex and energy dependent. The reduced mass is 
given by $\mu=M_Nm_K/(M_N+m_K)$. By starting from the effective interaction 
$V^{\text{eff}}(\sqrt{s})$ of Eq.~\eqref{eq:Veffective}, the ansatz for an 
equivalent local pseudopotential is 
\begin{equation}
    U(r,E) = \frac{g(r)}{2\,\tilde{\omega}}\frac{M_N}{\sqrt{s}} 
    \,V^{\text{eff}}(\sqrt{s}) ,
    \label{eq:potential1}
\end{equation}
with a form factor $g(r)$ representing the finite range of the interaction. 
The reduced energy $\tilde{\omega}$ is given by
\begin{align}
    \tilde{\omega}(\sqrt{s}) 
    & = \frac{\omega_K\,E_N}{ \omega_K + E_N} ,
    \nonumber
\end{align}
with
\begin{align}
    E_N 
    &= \frac{s-m_K^2+M_N^2}{2\sqrt{s}}   ,
    \quad
    \omega_K =\frac{s-M_N^2+m_K^2}{2\sqrt{s}} .
    \nonumber 
\end{align}
The energy $E$ appearing in the Schr\"odinger equation 
(\ref{eq:schroedinger}) is related to the total c.m. energy of the two-body 
system by
\begin{equation}
    E=\sqrt{s}-M_N-m_K .
    \nonumber
\end{equation}

For orientation, consider the zero-range $I=0$ $\bar{K}N$ pseudopotential 
generated by the leading Weinberg-Tomozawa term in the heavy-baryon limit 
$(M_N\rightarrow\infty)$; inserting $V^{\text{eff}} = V_{11}$ in 
Eq.~(\ref{eq:potential1}) one has
\begin{equation}
   U^{I=0}_{\text{WT}}(r) = - \frac{3}{ 4\,f^2}\,\delta^3(\vec{r}\,).
    \nonumber
\end{equation}
With $f\simeq 0.1$ GeV, the volume integral of this potential is 
$\int d^3r\, U_{\text{WT}} \simeq -0.58$ GeV.

The coupled-channel dynamics encoded in $\tilde{V}_{11}$ of 
Eq.~(\ref{eq:Veffective}) involves finite range effects through the 
$\pi\Sigma$ loops, which are given a minimal parametrization in terms of the
form factor $g(r)$. We choose a Gaussian ansatz
\begin{align}
    g(r) =& 
    \frac{1}{\pi^{3/2}b^3}\,e^{-r^2/b^2},
    \nonumber
\end{align}
with the range parameter $b$. This range parameter should reflect the 
subtraction constant $a_{\bar{K}N}$ used in the chiral coupled-channel 
approach.

\subsection{Comparison of scattering amplitudes}\label{subsec:amp}

The $\bar{K}N$ scattering amplitude $F_{\bar{K}N}$ derived from the 
potential $U(r,E)$ is now calculated in the usual way. The $s$-wave 
scattering amplitude is 
\begin{equation}
    F_{\bar{K}N} = \frac{1}{k(\cot \delta_0-i)} ,
    \nonumber 
\end{equation} 
where the phase shift $\delta_0$ is determined by the asymptotic wave 
function, 
\begin{equation}
    \frac{u(r)}{r} \to A_0[\cos \delta_0 j_0(kr) 
    -\sin \delta_0 n_0(kr)]
    \quad \text{for} \quad r\to \infty ,
    \nonumber
\end{equation}
with spherical Bessel and Neumann functions $j_0$ and $n_0$. The wave number
$k=\sqrt{2\mu E}$ becomes imaginary below threshold, $E<0$. 
 
Given $V^{\text{eff}}(\sqrt{s})$ as input, the range parameter $b$ is then 
fixed by requiring that the real part of the $\bar{K}N$ amplitude develops 
its zero at $\sqrt{s} \simeq 1420$ MeV to satisfy the condition for the 
quasibound $\bar{K}N$ state at this point. For the HNJH model, this 
condition determines $b = 0.47$ fm. Note that this scale is somewhat smaller
than the typical range associated with vector meson exchange, the picture 
that one has in mind as underlying the vector current interaction generating
the Weinberg-Tomozawa term.
 
\begin{figure*}[tbp]
    \centering
    \includegraphics[width=0.75\textwidth,clip]{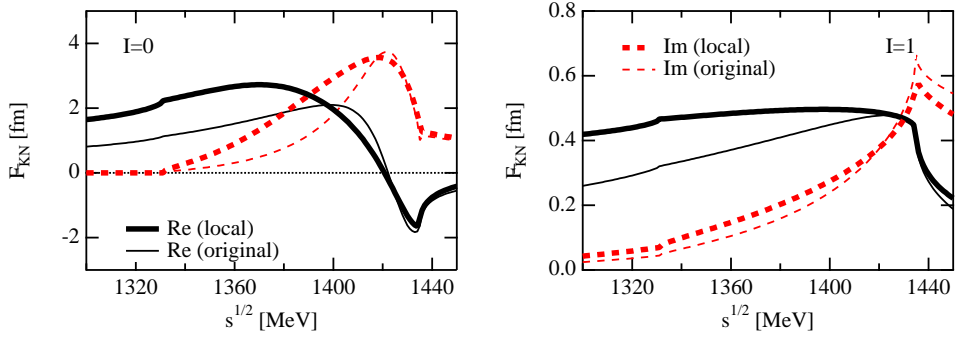}
    \caption{\label{fig:potential1}
    (Color online) Scattering amplitudes $F_{\bar{K}N}$ from the local 
    potential $U(r,E)$ (thick lines) and from the amplitude $T^{\text{eff}}$
    in the original chiral coupled-channel approach (thin lines) obtained by
    using the HNJH model for the $I=0$ channel (left) and the $I=1$ channel 
    (right). Real parts are shown as solid lines and imaginary parts as 
    dashed lines.}
\end{figure*}

With $b = 0.47$ fm fixed, the $I=0$ and $I=1$ amplitudes generated by the 
equivalent local pseudopotential $U(r,E)$ reproduce the full $\bar{K}N$ 
coupled-channel amplitudes perfectly well in the threshold and subthreshold 
region above $\sqrt{s} \simeq 1420$ MeV. However, at energies below the 
quasibound state, the local ansatz [Eq.~(\ref{eq:potential1})] does not 
extrapolate correctly into the far-subthreshold region. One has to keep in 
mind that the complex, off-shell effective $\bar{K}N$ interaction is in 
general nonlocal and energy dependent to start with. Its detailed behavior 
over a broader energy range cannot be approximated by a simple local 
potential without paying the price of extra energy dependence. This is 
demonstrated in Fig. \ref{fig:potential1}. In the subthreshold region below 
$\sqrt{s}<1400$ MeV, the amplitudes calculated with the local potential 
overestimate the ones resulting from the coupled-channel approach 
significantly, in both $I=0$ and $I=1$ channels. One observes that 
subthreshold extrapolations using a naive local potential tend to give much 
stronger $\bar{K}N$ attraction than what chiral coupled-channel dynamics 
actually predicts. Corrections to the energy dependence of the local 
potential need to be applied to repair this deficiency.  
   
\subsection{Improved local potentials and \\ uncertainty analysis} 
\label{subsec:correction}

\begin{figure*}[tbp]
    \centering
    \includegraphics[width=0.75\textwidth,clip]{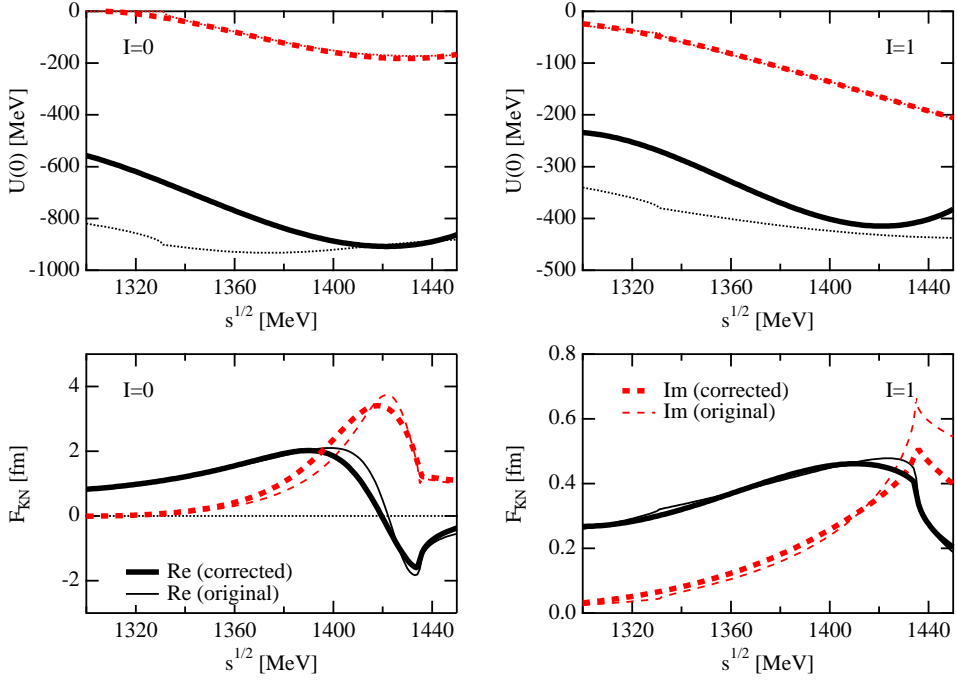}
    \caption{\label{fig:potential2}
    (Color online) Upper panels: Strength of the fitted potential at $r=0$ 
    (thick lines) and the strength without correction 
    [Eq.~\eqref{eq:potential1}; dotted lines] with the HNJH model. Lower 
    panels: Scattering amplitude $f$ from the local potential (thick lines) 
    and the amplitude $T_{\text{eff.}}$ in the original chiral unitary 
    approach (thin lines) with the HNJH model. The real parts are shown by 
    the solid lines and the imaginary parts are depicted by the dotted 
    lines. Left: $I=0$ channel. Right: $I=1$ channel.}
\end{figure*}

The necessary corrections just mentioned can easily be implemented by 
introducing a third-order polynomial in $\sqrt{s}$,
\begin{align}
    &U(r=0,E)
    = K_0+ K_1\sqrt{s} + K_2(\sqrt{s})^2+ K_3(\sqrt{s})^3 ,
    \nonumber \\
    &\quad \quad 1300 \leq \sqrt{s} \leq 1450\text{ MeV} ,
    \nonumber
\end{align}
to reproduce the full coupled-channel result for $F_{\bar{K}N}$ at
$\sqrt{s}<1400$ MeV. The coefficients $K_i$ are summarized in 
Tables~\ref{tbl:coefficientsI0} and \ref{tbl:coefficientsI1} (rows of the 
HNJH model). The strength of the fitted potential at $r=0$ is shown in the 
upper panel of Fig.~\ref{fig:potential2} by thick lines. The strength of the
potential in its original form [Eq.~\eqref{eq:potential1}] is also shown by 
the dotted lines. 

The resulting scattering amplitudes are presented in the lower panel of 
Fig.~\ref{fig:potential2} together with the amplitudes resulting from the 
chiral unitary approach. As seen in the figure, a 20\% reduction of the 
local potential is required to match the amplitudes in the energy region 
$\sqrt{s}<1400$ MeV. As a consequence the energy dependence of the local 
potential becomes stronger. The matching of the amplitude in the energy 
range around the quasibound state and close to threshold remains unchanged,
keeping the range parameter $b$ at its previously determined value. With its
improved energy dependence, the overall matching of the ``equivalent" local 
$\bar{K}N$ potential is now quite satisfactory, but the attraction in the 
far-subthreshold region is substantially weaker than naively anticipated.

To estimate again possible uncertainties, we apply the corrections
to the energy dependence of the potentials obtained with all four variants 
of the chiral models studied in Sec.~\ref{subsec:theoretical}. The range 
parameters are determined so as to reproduce the position of the quasibound 
$\bar{K}N$ state at the same energy $\sqrt{s} \simeq 1420$ MeV as found with
the coupled-channel amplitudes. The resulting values of the range parameter
$b$ are summarized in Table~\ref{tbl:range}. The expected reciprocal 
relationship between subtraction constants $a_{\bar{K}N}$ and Gaussian range
parameters is evident.

\begin{table}[tbp]
    \centering
    \caption{Range parameters $b$ of the local potential and the subtraction
    constants $a_{\bar{K}N}$ used in the chiral unitary approach.}    
    \begin{ruledtabular}
        \begin{tabular}{lcc}
        Reference & $b$ (fm) & $a_{\bar{K}N}$  \\
        \hline
        ORB~\cite{Oset:2001cn} & 0.52 & $-1.84$ \\
        HNJH~\cite{Hyodo:2002pk} & 0.47 & $-1.96$ \\
        BNW~\cite{Borasoy:2005ie} & 0.51 & $-1.86$ \\
        BMN~\cite{Borasoy:2006sr} & 0.41 & $-2.21$ \\
    \end{tabular}
    \end{ruledtabular}
    \label{tbl:range}
\end{table}

\begin{table*}[tbp]
    \centering
    \caption{Coefficients of the polynomial of the effective 
    interaction for $I=0$.}    
    \begin{ruledtabular}
        \begin{tabular}{lllllllll}
        Reference &
	\multicolumn{2}{c}{$K_0$ ($10^5$MeV)}
	& \multicolumn{2}{c}{$K_1$ ($10^2$MeV$^0$)}
	& \multicolumn{2}{c}{$K_2$ ($10^{-1}$MeV$^{-1}$)}
	& \multicolumn{2}{c}{$K_3$ ($10^{-4}$MeV$^{-2}$)}\\
        & $\re $ & $\im $ 
        & $\re $ & $\im $ 
        & $\re $ & $\im $ 
        & $\re $ & $\im $  \\
        \hline
        ORB~\cite{Oset:2001cn}
	& $-3.9321$ & $-4.5613$ 
	& $\phantom{0}8.9088$ & $10.097$ 
	& $-\phantom{0}6.709$ & $-\phantom{0}7.4364$ 
	& $1.6771$ & $1.8219$ \\
        HNJH~\cite{Hyodo:2002pk} 
	& $-5.1020$ & $-4.3660$ 
	& $11.453$ & $\phantom{0}9.6378$ 
	& $-\phantom{0}8.5527$ & $-\phantom{0}7.0773$ 
	& $2.1218$ & $1.7285$ \\
        BNW~\cite{Borasoy:2005ie} 
	& $-4.3330$ & $-6.6603$ 
	& $\phantom{0}9.8635$ & $14.710$ 
	& $-\phantom{0}7.4619$ & $-10.812$ 
	& $1.8738$ & $2.6443$ \\
        BMN~\cite{Borasoy:2006sr}	
	& $-6.6455$ & $-4.0390$ 
	& $14.873$ & $\phantom{0}8.8408$ 
	& $-11.075$ & $-\phantom{0}6.4345$ 
	& $2.7401$ & $1.5568$ 
    \end{tabular}
    \end{ruledtabular}
    \label{tbl:coefficientsI0}
\end{table*}
\begin{table*}[tbp]
    \centering
    \caption{Coefficients of the polynomial of the effective 
    interaction for $I=1$.}    
    \begin{ruledtabular}
        \begin{tabular}{lllllllll}
        Reference &
	\multicolumn{2}{c}{$K_0$ ($10^5$MeV)}
	& \multicolumn{2}{c}{$K_1$ ($10^2$MeV$^0$)}
	& \multicolumn{2}{c}{$K_2$ ($10^{-1}$MeV$^{-1}$)}
	& \multicolumn{2}{c}{$K_3$ ($10^{-4}$MeV$^{-2}$)}\\
        & $\re $ & $\im $ 
        & $\re $ & $\im $ 
        & $\re $ & $\im $ 
        & $\re $ & $\im $  \\
        \hline
        ORB~\cite{Oset:2001cn}
	& $-6.2984$ & $-0.63191$ 
	& $13.939$ & $1.3709$ 
	& $-10.272$ & $-0.98412$ 
	& $2.5195$ & $0.23337$ \\
        HNJH~\cite{Hyodo:2002pk} 
	& $-4.4348$ & $-0.67630$ 
	& $\phantom{0}9.8340$ & $1.4675$ 
	& $-\phantom{0}7.2582$ & $-1.0532$ 
	& $1.7818$ & $0.24953$ \\
        BNW~\cite{Borasoy:2005ie} 
	& $-2.6295$ & $-0.48818$ 
	& $\phantom{0}5.8297$ & $1.0484$ 
	& $-\phantom{0}4.2999$ & $-0.74387$ 
	& $1.0542$ & $0.17396$ \\
        BMN~\cite{Borasoy:2006sr}	
	& $-7.5894$ & $-0.52306$ 
	& $16.797$ & $1.1074$ 
	& $-12.375$ & $-0.77213$ 
	& $3.0330$ & $0.17669$ 
    \end{tabular}
    \end{ruledtabular}
    \label{tbl:coefficientsI1}
\end{table*}

In all cases studied, the local potential produces scattering amplitudes 
that are systematically too large (i.e., too strongly attractive) at the 
lower energy side, when compared with the amplitudes of the chiral 
coupled-channel approach. Repeating the procedures as before, we improve the
potentials, correcting the far-subthreshold energy dependence by matching 
the scattering amplitudes to the chiral coupled-channel results. These 
corrections are again performed by adjusting the third-order polynomial 
[Eq.~(IV C)] in each case. The coefficients of the polynomial fit are 
collected in Tables~\ref{tbl:coefficientsI0} and \ref{tbl:coefficientsI1}. 

\begin{figure*}[tbp]
    \centering
    \includegraphics[width=0.75\textwidth,clip]{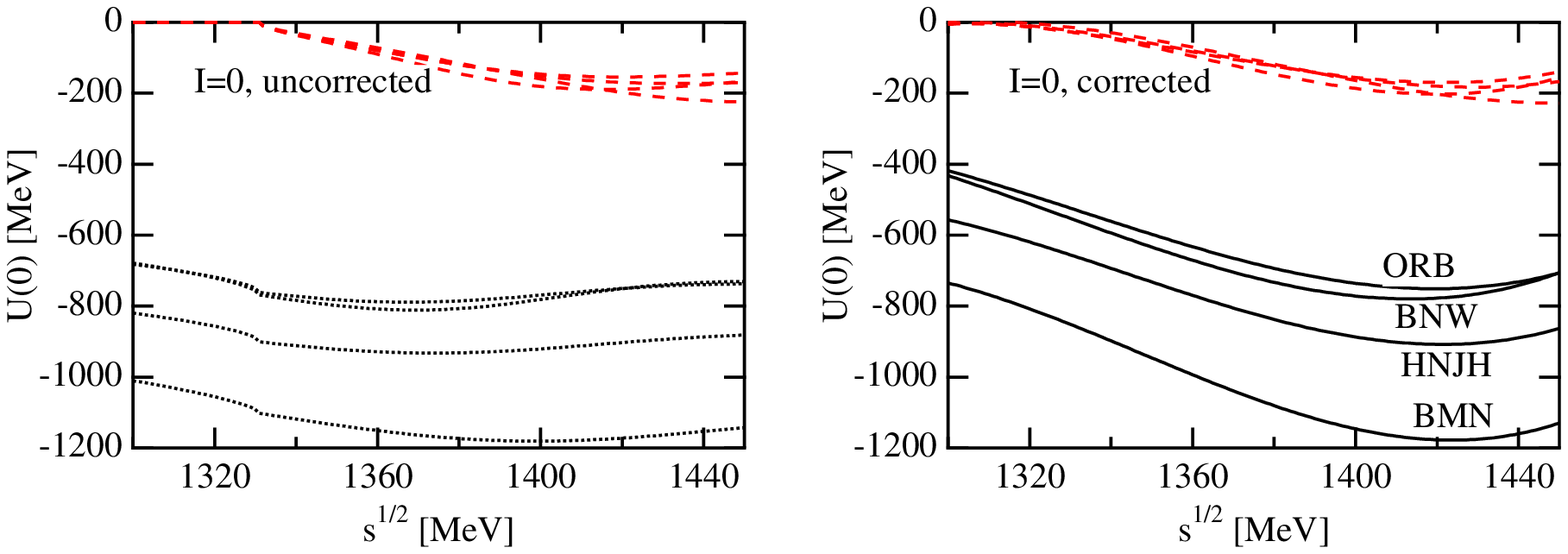}
    \caption{\label{fig:potential3_1}
    (Color online) Strength of the local $I = 0$ $\bar{K}N$ potential 
    $U(r,E)$ at $r=0$ obtained by using different models as explained in the
    text. Left: Uncorrected potentials, with real parts shown as dotted 
    lines. Right: Corrected potentials, with real parts shown as solid 
    lines. Imaginary parts are depicted as dashed lines. The ordering of 
    model assignments is the same in both panels.}
\end{figure*}

\begin{figure*}[tbp]
    \centering
    \includegraphics[width=0.75\textwidth,clip]{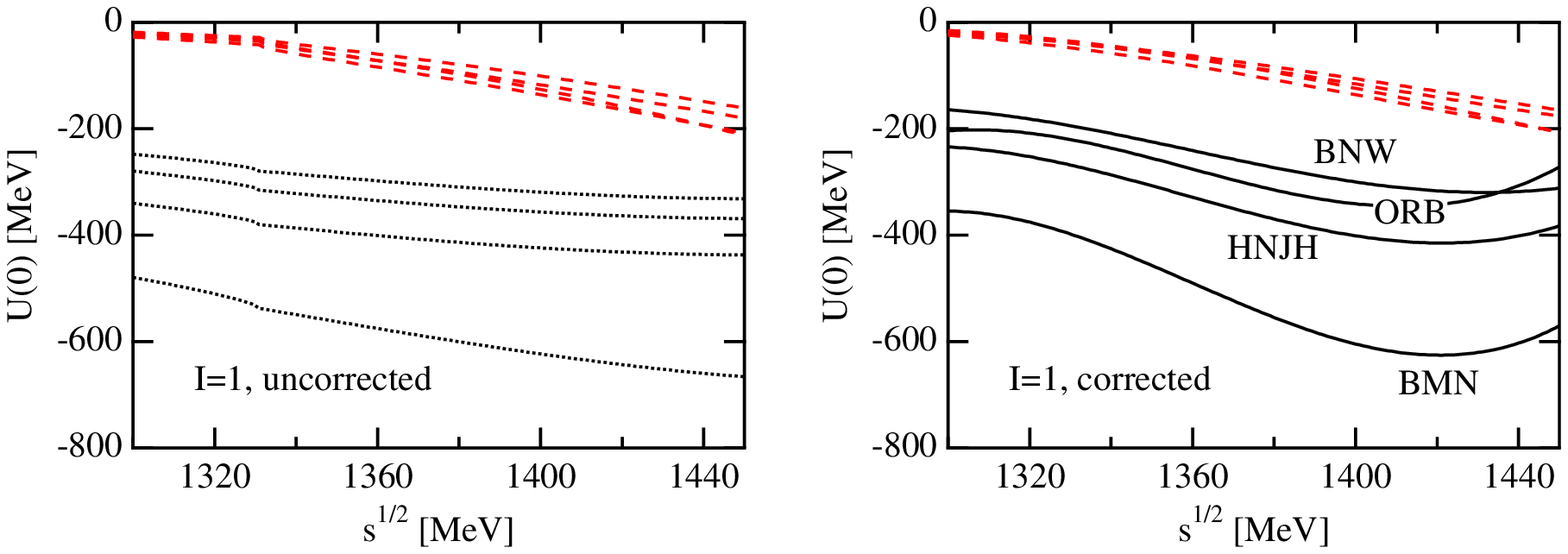}
    \caption{\label{fig:potential3_2}
    (Color online) Strength of the local $I = 1$ $\bar{K}N$ potential 
    $U(r,E)$ at $r=0$ obtained by using different models as explained in the
    text. Left: Uncorrected potentials, with real parts shown as dotted 
    lines. Right: Corrected potentials, with real parts shown as solid 
    lines. Imaginary parts are depicted as dashed lines. The ordering of 
    model assignments is the same in both panels.}
\end{figure*}

\begin{figure*}[tbp]
    \centering
    \includegraphics[width=0.75\textwidth,clip]{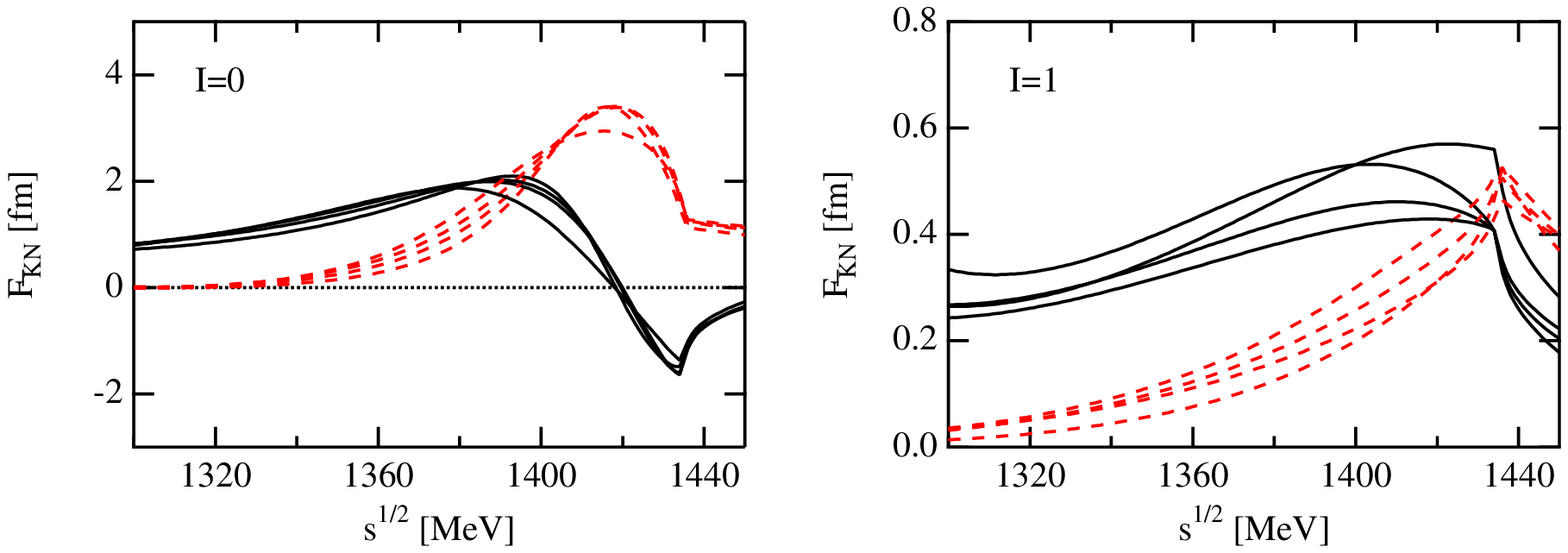}
    \caption{\label{fig:potential3_3}
    (Color online)  Scattering amplitudes $F_{\bar{K}N}$ from the corrected 
    local potentials $U(r,E)$ for the four different models referred to in 
    Figs.~\ref{fig:potential3_1} and \ref{fig:potential3_2}. Real parts are 
    shown as solid lines and imaginary parts as dashed lines. Left: $I=0$ 
    channel. Right: $I=1$ channel.}
\end{figure*}

The comparison between uncorrected and corrected potential strengths at the
origin, $U(r=0,E)$, is shown in Figs.~\ref{fig:potential3_1} and 
\ref{fig:potential3_2}. The resulting scattering amplitudes are presented in
Fig.~\ref{fig:potential3_3}. Note that all the potentials pictured in the 
right-hand panels of Figs. \ref{fig:potential3_1} and \ref{fig:potential3_2}
are ``equivalent'' in that they reproduce the same scattering amplitudes 
both 
on- and off-shell (below threshold) to good approximation. The differences 
in the potential strengths at $r=0$ are largely balanced by differences in 
the range parameters $b$, such that the volume integrals $\int d^3r\,U$ are 
in a reduced band (illustrated by the smaller spread of $V^{\text{eff}}$ in 
Fig.~\ref{fig:Effective_model}) for all different versions of the chiral 
effective $\bar{K}N$ interactions.

\subsection{Comparison with the\\ 
phenomenological AY potential}\label{subsec:AYpotential}

Finally, let us compare our results with the phenomenological 
Akaishi-Yamazaki (AY) potential~\cite{Akaishi:2002bg, Yamazaki:2007cs}. This
potential was introduced on purely phenomenological grounds by fitting the 
$\bar{K}N$ scattering data and the PDG value of the $\Lambda(1405)$ 
resonance. In $\bar{K}N$-$\pi\Sigma$ coupled channels with $I=0$, the AY 
potential reads
\begin{equation}
    v_{ij}(r)
    =\begin{pmatrix}
       -436 & -412 \\
         -412 & 0 & 
    \end{pmatrix}
    \exp [-(r/b)^2] \quad \text{[MeV]}
    \nonumber ,
\end{equation}
with $b\sim 0.66$ fm. Apart from the missing energy dependence, there are 
further qualitative differences between this potential and the interaction 
based on chiral SU(3) dynamics. The most drastic difference is the absence 
of a direct  $\pi\Sigma\to \pi\Sigma$ coupling in the phenomenological 
potential. This is in sharp contrast with chiral dynamics, since the 
attractive interaction in the diagonal $\pi\Sigma$ channel is sufficiently 
strong to generate a resonance, as we have discussed in the previous 
sections. The coupled-channel dynamics leads to the quasibound structure in 
the $\bar{K}N$ system at around 1420 MeV. 

The phenomenological AY model~\cite{Yamazaki:2007cs} starts from the 
``ansatz'' that the $\Lambda(1405)$ resonance is a $K^-p$ bound state. It 
shares this principal feature with chiral dynamics. However, the absence of 
the $\pi\Sigma\to \pi\Sigma$ coupling in the AY model has as its consequence
that the $\Lambda(1405)$ is represented by only a single pole and the mass 
spectrum in the $\pi\Sigma$ channel is identical to that in the $\bar{K}N$ 
channel. This incorrectly implies a $\bar{K}N$ single-channel interaction 
that is too strongly attractive.

\begin{figure*}[tbp]
    \centering
    \includegraphics[width=0.75\textwidth,clip]{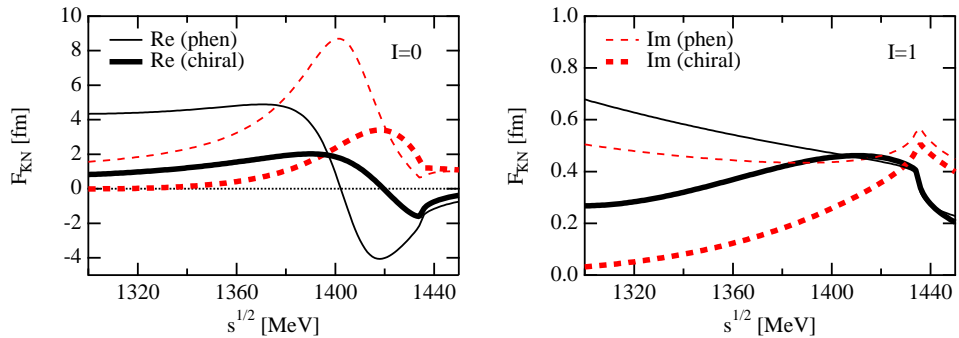}
    \caption{\label{fig:potential_AY}
    (Color online) Comparison of the scattering amplitudes with 
    phenomenological potential. Thick lines denote the results of the 
    potentials derived in this work and thin lines denote the results of 
    the phenomenological potential~\cite{Yamazaki:2007cs}. The real parts 
    are shown by solid lines and the imaginary parts are depicted 
    by dotted lines. Left: $I=0$ channel. Right: $I=1$ channel.}
\end{figure*}

In Ref.~\cite{Yamazaki:2007cs}, the equivalent single-channel potentials are
given (in MeV) as
\begin{align}
    v_{\bar{K}N}^{I=0}(r)
    =&\,(-595-i\, 83)\, \exp [-(r/0.66\,\text{fm})^2], \nonumber \\
    v_{\bar{K}N}^{I=1}(r)
    =&\,(-175-i\, 105)\, \exp [-(r/0.66\,\text{fm})^2] .
    \label{eq:AYpotential}
\end{align}
The amplitudes resulting from these potentials are shown in 
Fig.~\ref{fig:potential_AY} by thin lines, to be compared with our present 
results (thick lines). The behavior of the amplitudes derived from the 
phenomenological potentials is seen to be drastically different from the 
chiral dynamics prediction, especially in the lower subthreshold energy 
region. The difference is certainly beyond the theoretical uncertainties 
estimated in Sec.~\ref{subsec:theoretical}. As we discussed, chiral dynamics
locates the $\Lambda(1405)$ in the $\bar{K}N$ amplitude around 1420 MeV, 
whereas the phenomenological AY potential uses the PDG value of around 1405 
MeV. The difference is not only in the position of the $\Lambda(1405)$ but 
also in the magnitude of the amplitude. Note also that the imaginary parts 
remain finite even below the $\pi\Sigma$ threshold, since there is no energy
dependence in the potentials~\eqref{eq:AYpotential}.

However, above the $\bar{K}N$ threshold, both chiral and phenomenological 
amplitudes behave similarly as both approaches are adjusted to describe 
existing data. These and the previous observations clearly point out the 
considerable ambiguities involved in the extrapolation of the amplitudes 
below the $\bar{K}N$ threshold. Existing data sets are not sufficient to 
constrain the $\bar{K}N$ interaction in the energy region relevant for 
subthreshold antikaon-nucleon physics. At this point the minimal theoretical
constraints from chiral SU(3) dynamics turn out to be of crucial importance.

\section{Summary}\label{sec:summary}

We have constructed an effective $\bar{K}N$ potential in coordinate space 
based on chiral SU(3) dynamics. This procedure involves two steps:
\begin{enumerate}
    \item  transforming the coupled-channel dynamics into a single 
    channel $\bar{K}N$ interaction and
    \item  translating this effective interaction to an ``equivalent'' 
    local potential.
\end{enumerate}
Step 1 is exact within the chiral coupled-channels approach, whereas step 2
involves approximations. 

In performing step 1 we have systematically investigated how the dynamics
of the $\pi\Sigma$ channel (and of other channels, which turn out not to be 
important) influence the effective $\bar{K}N$ interaction. It is found that 
the $\pi\Sigma$ interaction generates a broad resonance whereas the 
$\bar{K}N$ interaction produces a weakly bound state. The attractive forces 
in the $\bar{K}N$ and $\pi\Sigma$ coupled channels cooperate to form the
$\Lambda(1405)$ as a $\bar{K}N$ quasibound state embedded in the $\pi\Sigma$
continuum. 

As a consequence of the two-pole structure of the coupled 
$\bar{K}N$-$\pi\Sigma$ system, the $\bar{K}N$ quasibound state is located at
$\sqrt{s}\simeq1420$ MeV, at a mass shifted upward from the PDG value of 
$\Lambda(1405)$ deduced from the maximum of the $\pi\Sigma$ mass spectrum. 

We have examined several versions of chiral SU(3) dynamical models to 
estimate theoretical uncertainties. The occurrence of the quasibound state 
at $\sqrt{s}\simeq1420$ MeV turns out to be model independent, irrespective 
of the wide spread in the locations of the second pole. This implies that 
the $\bar{K}N$ binding energy associated with the $\Lambda(1405)$ is 
actually not 27 MeV but only 12 MeV, indicating significantly less 
attraction in the effective $\bar{K}N$ interaction than previously 
anticipated on purely phenomenological grounds.

In step 2, we have constructed an equivalent local potential, with a 
Gaussian r-space form factor reflecting finite range effects. The range 
parameter is adjusted so as to reproduce the position of the $\bar{K}N$ 
quasibound state and found to be somewhat smaller than that expected from a 
vector meson exchange picture. In any case, a local parametrization of the 
$\bar{K}N$ effective interaction works only with an adjustment of its energy
dependence. This extra energy dependence considerably reduces the attractive
strength of the potential in the subthreshold region as compared to naive 
expectations.

Uncertainties concerning subthreshold extrapolations of the effective 
$\bar{K}N$ interaction still remain as long as the constraints from 
threshold $\bar{K}N$ data and from the $\pi\Sigma$ mass spectrum in the 
$I=0$ channel are relatively weak and partly ambiguous. There is a strong 
demand for further improvements in this empirical data base. Nonetheless, 
the theoretical constraints on the strengths of the $\bar{K}N$ and 
$\pi\Sigma$ interactions from chiral SU(3) dynamics are certainly mandatory
to reduce the freedom of extrapolation.

The potential so obtained suggests itself for applications in 
antikaon-nuclear few-body calculations. Results for the $K^-pp$ prototype 
system are reported in Ref.~\cite{Dotesan}. The overall picture presented 
here, constrained by chiral SU(3) dynamics, differs strongly from 
previous purely phenomenological approaches, in that the resulting effective
potential is significantly less attractive in the energy range relevant to 
the discussion of deeply bound antikaon-nuclear clusters.  

\begin{acknowledgments}
    The authors thank Avraham Gal and Akinobu Dot\'e for many helpful 
    discussions. This project is partially supported by BMBF, GSI, and by 
    the DFG excellence cluster ``Origin and Structure of the Universe." 
    T.~H. thanks the Japan Society for the Promotion of Science (JSPS) for 
    financial support. His work is supported in part by the Grant for 
    Scientific Research (No.\ 19853500) from the Ministry of Education, 
    Culture, Sports, Science and Technology (MEXT) of Japan. This research 
    is part of the Yukawa International Program for Quark-Hadron Science. 
\end{acknowledgments}

\end{document}